\documentclass[a4paper,11pt]{article}
\pdfoutput=1 
\usepackage{jinstpub} 
\usepackage{lineno}
\graphicspath{./}
\usepackage{mathtools}
\DeclareMathOperator{\atan2}{atan2}

\title{Electron beam polarimeter and energy spectrometer}
\author[a,b]{N. Yu. Muchnoi%
\note{Corresponding author.}}
\affiliation[a]{Budker Institute of Nuclear Physics, Lavrentieva 11, Novosibirsk, Russian Fedration}
\affiliation[b]{Novosibirsk State University, Pirogova 2, Novosibirsk, Russian Federation}

\emailAdd{N.Yu.Muchnoi@inp.nsk.su}

\abstract{The backscattering of laser radiation by a relativistic electron beam is a widely used method for measuring the spin polarization of electrons. 
We consider again the properties of the scattering process paying special attention to recoil electrons. 
Based on this consideration we propose the concept of the Compton polarimeter in which, in addition to all the polarization components, it becomes possible to accurately measure the energy and other parameters of the electron beam. 
To demonstrate the capabilities of the method we conduct a Monte Carlo simulations of the polarimeter developed for the FCC-ee project.}

\keywords{Accelerator Subsystems and Technologies, Beam-line instrumentation (beam position and profile monitors, beam-intensity monitors, bunch length monitors), Instrumentation for particle accelerators and storage rings - high energy (linear accelerators, synchrotrons).}

\begin{document}

\maketitle

\section{Inverse Compton scattering}\label{section1}

Shortly after the invention of lasers it was pointed out in refs.~\cite{milburn_electron_1963,arutyunian_compton_1963} that Inverse Compton Scattering (ICS) of laser light by relativistic electron beam would produce useful yields of high energy polarized photons. 
In 1969 such a beam had been realized at SLAC for a study of $\gamma p$ interactions~\cite{ballam_total_1969}.
The effect of self-polarization of relativistic electrons due to emission of the spin-flip synchrotron radiation was predicted in ref.~\cite{sokolov_polarization_1964} and observed at the first ever $e^+e^-$ collider ACO~\cite{belbeoch_r_new_1968} due to polarization effects in intra-beam scattering.
Soon the existence of this phenomenon was confirmed on the storage rings VEPP-2 and SPEAR as described in review~\cite{derbenev_radiative_1978}.
Theoretical and experimental studies of spin dynamics in storage rings made it possible to develop a method of resonant depolarization described e. g. in ref.~\cite{skrinskii_precision_1989} and references therein. 
For the first time it was used for the beam energy calibration in experiments on precise determination of the $\Phi$ and $K$ mesons masses~\cite{sidorov_results_1976}.
At higher beam energies in addition to other drawbacks the intra-beam scattering ceases to be a good tool for polarization measurement.
ICS was proposed in ref.~\cite{bayer_v_n_determination_1969} as one of the processes allowing to measure the transverse spin polarization of high-energy electron beam. 
At the spin polarimeters of the VEPP-4 \cite{kezerashvili_colliding_1992} and VEPP-2M \cite{artamonov_high_1982} colliders the synchrotron or undulator radiation was scattered in the $e^+e^-$ collision area.
ICS of laser light for transverse beam polarization measurement was first used at SPEAR~\cite{gustavson_back_1979} and then at circular colliders PETRA~\cite{bremer_petra-polarimeter_1981}, DORIS-II~\cite{barber_precision_1984}, LEP~\cite{knudsen_first_1991,arnaudon_measurement_1992}, HERA~\cite{barber_hera_1993} and VEPP-4M~\cite{blinov_status_2020}.
The ICS for longitudinal spin polarimetry was realized at SLAC linear collider~\cite{woods_m_scanning_1996}, AmPS ring at NIKHEF~\cite{passchier_compton_1998}, HERA collider at DESY~\cite{beckmann_longitudinal_2002} and Jlab~\cite{escoffier_accurate_2005}.
The ICS polarimeters are also planned for future collider experiments ILC~\cite{bartels_precision_2010,telnov_beam_2003}, FCC-ee~\cite{blondel_polarization_2019}, CEPC~\cite{nikitin_opportunities_2019,xia_cepc_2021} and EIC~\cite{accardi_electron-ion_2016,skaritka_conceptual_2018}.

\pagebreak

\begin{figure}[h]
\centering
\includegraphics[width=\linewidth]{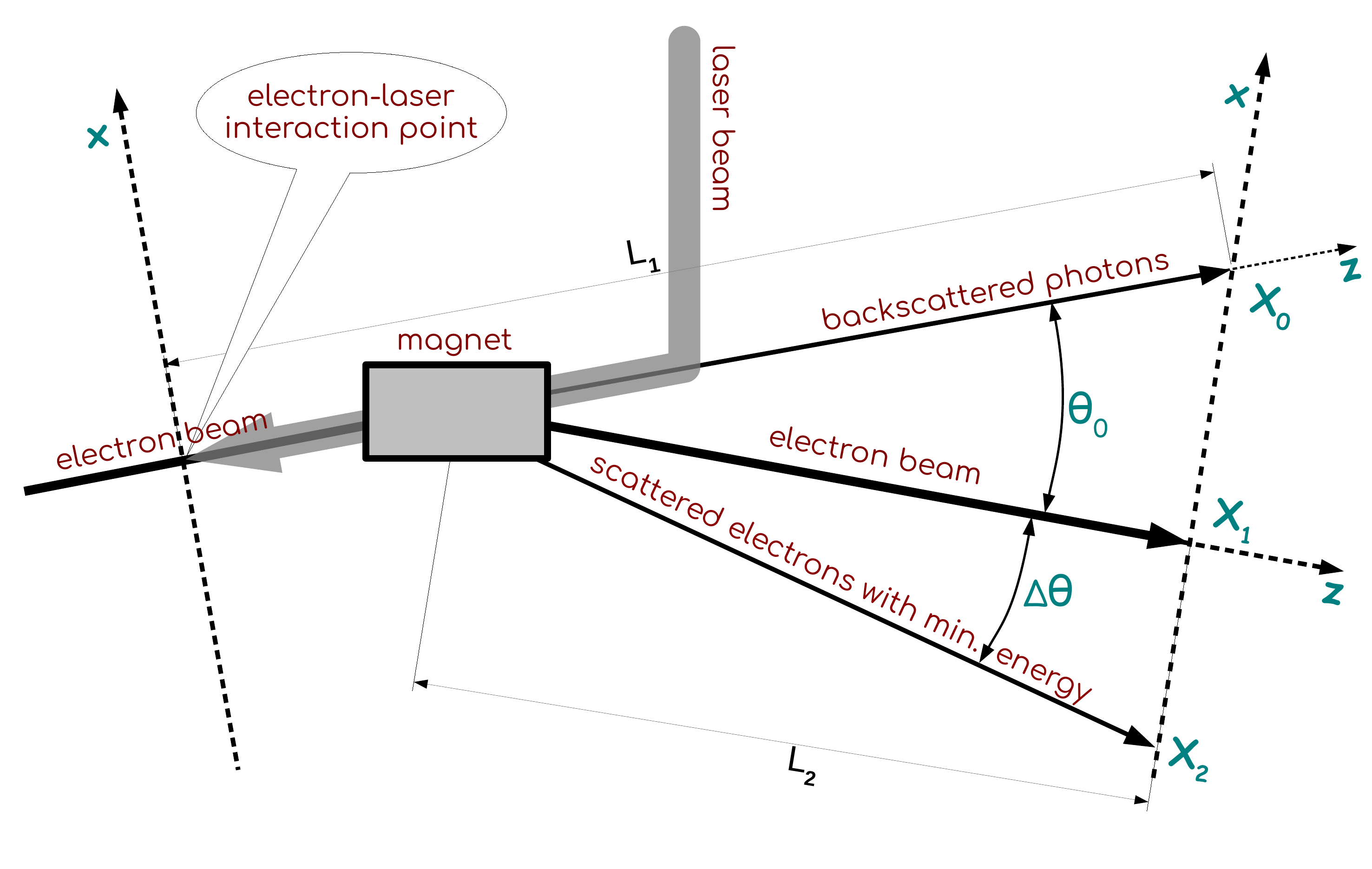}
\caption{Layout of ICS experiment.}
\label{fig:cxema}
\end{figure}

The practical realization of the ICS looks like it is shown in figure~\ref{fig:cxema}.
Laser radiation is inserted into the accelerator vacuum chamber and focused to the interaction point.
The plane of the figure is the plane of machine, the $z$-axis of the curvilinear coordinate system is always directed along the electron beam momentum.
The $x$-axis is shown in figure~\ref{fig:cxema} while the $y$-axis is perpendicular to the plane of machine.
The dipole magnet separates the scattered photons from the electron beam.
The beam is bent by the angle $\theta_0$ and the recoil electrons are deflected by a larger angle $\theta_0+\Delta\theta$. 

Angle $\theta_0$ depends on the ratio of the bending field integral to the electron energy $\varepsilon_0=\gamma m c^2$:
\begin{equation}
\theta_0 = \displaystyle\frac{\int \!\! B_\perp dl}{\gamma(mc/e)}, \text{~~where~~} \frac{mc}{e} = 1.704509024\cdot10^{-3} [\text{T~m}].
\label{bend}
\end{equation}

Consider the scattering of a low energy photon with wavelength $\lambda_0$ and energy $\hbar\omega_0$ by an electron with energy $\varepsilon_0$. 
Denoting the energies of the photon and electron after scattering as $\hbar\omega$ and $\varepsilon$, their ratio is the dimensionless parameter $u=\hbar\omega/\varepsilon$ as in refs.~\cite{herrmann_compton_1972,abakumova_backscattering_2013}.

When $\gamma\gg1$ one can neglect $\hbar\omega_0 \ll \varepsilon_0, \varepsilon, \hbar\omega$ in energy conservation equation and thus
\begin{equation}
u
\equiv\displaystyle\frac{\hbar\omega}{\varepsilon}
=\displaystyle\frac{\hbar\omega}{\varepsilon_0-\hbar\omega}
=\displaystyle\frac{\varepsilon_0-\varepsilon}{\varepsilon}
=\displaystyle\frac{\theta_e}{\theta_\gamma},
\label{u}
\end{equation}
where $\theta_\gamma$ and $\theta_e$ are the scattering angles of photon and electron with respect to the initial electron momentum.
The equality $\varepsilon \theta_e = \hbar \omega \theta_\gamma$ follows from the law of conservation of the transverse momentum of the system. 

Parameter $u$ lies in the range $u \in \left[0,\kappa\right]$, where
 $\kappa$ is twice the initial energy of the photon in the rest frame of the electron, expressed in units of the electron rest energy:
\begin{equation}
\kappa = 2 \cdot \frac{ 2\gamma \hbar\omega_0}{mc^2} = \frac{4\hbar\omega_0\varepsilon_0}{(mc^2)^2}.
\label{kappa}
\end{equation}
If $\alpha\ne\pi$ but $\tan(\alpha/2) \gg 1/\gamma$, $\kappa \sin^2\!\left(\alpha/2\right)$ should be used instead of $\kappa$.
Table~\ref{scattering_parameters} shows how the parameters of scattered photon and recoil electron are derived from $u$ and $\kappa$ variables.
\begin{table}[h]
\label{scattering_parameters}
\caption{Parameters of scattered particles.}
\begin{equation*}
\boxed{
\begin{aligned}
                           & &\text{scattered photon:}                             & & \text{recoil electron:}  \\
\text{Energy:}             & &\hbar\omega   = \frac{\varepsilon_0 u}{1+u}          &,& \varepsilon =       \frac{\varepsilon_0}{1+u}. \\
\text{Edge energy:}        & &\hbar\omega_{max}  = \frac{\varepsilon_0\kappa}{1+\kappa} &,& \varepsilon_{min} = \frac{\varepsilon_0}{1+\kappa}. \\
\text{Scattering angle:}   & &\theta_\gamma = \frac{1}{\gamma}\sqrt{\kappa/u-1}    &,& \theta_e = \frac{u}{\gamma}\sqrt{\kappa/u-1}. \\
\text{Angle from bend:}    & &                \theta_0                            &,& u\theta_0. \\
\text{Maximum bend:}       & &                \theta_0                            &,& \Delta\theta=\kappa\theta_0. \\
\text{Azimuthal angle:}    & &\varphi                                              &,& \varphi+\pi. \\
\text{Horizontal angle:}   & &\eta_x=\theta_\gamma\cos\varphi-\theta_0         &,& \theta_x = u\theta_0 - \theta_e\cos\varphi. \\
\text{Vertical angle:}     & &\eta_y=\theta_\gamma\sin\varphi\;\;\;\;\;\;\;\; &,& \theta_y = \;\;\;\;\;\; -\;\theta_e\sin\varphi. \\
\end{aligned}}
\end{equation*}
\end{table}

Maximum scattering angle of a recoil electron $\max(\theta_e) = 2\hbar\omega_0/mc^2$ does not depend on the initial electron energy $\varepsilon_0$ and happens when  $u=\kappa/2$.
Since the energy of recoil electron is $\varepsilon=\varepsilon_0/(1+u)$, due to eq.~(\ref{bend}) such an electron will be bent by the magnet to the angle $\theta_0(1+u)$. 
According to figure~\ref{fig:cxema}, the first term $\theta_0$ defines the coordinate system rotation and is denoted in table~\ref{scattering_parameters} as the ``angle from bend'' and ``maximum bend'' for scattered photon.
The second term $u\theta_0$ defines the bend of the recoil electron due to its energy loss.
This is the ``angle from bend'' of recoil electron defined in table~\ref{scattering_parameters}.
The ``maximum bend'' of recoil electron $\Delta\theta = \kappa\theta_0$ happens when $u=\kappa$ and $\theta_\gamma=\theta_e=0$.
So far as $\Delta\theta/\theta_0 = \kappa$ the measurement of the ratio of angles $\Delta\theta$ to $\theta_0$ provides the possibility to determine the average energy of electrons in the beam according to the definition of $\kappa$ given in eq.~(\ref{kappa}):\\
\begin{equation}
\varepsilon_0 =  \frac{(mc^2)^2}{4\hbar\omega_0}\frac{\Delta\theta}{\theta_0}.
\label{yesenergy}
\end{equation}
This approach was proposed in ref.~\cite{muchnoi_ilc_2009}.
As it follows from eq.~(\ref{bend}), $\Delta\theta$ does not depend on $\varepsilon_0$: 
\begin{equation}
\Delta\theta = \frac{4\hbar\omega_0}{mc^2}\frac{\int \!\! B_\perp dl}{mc/e}.
\label{noenergy}
\end{equation}
Since $u$ is just a fractional part of $\kappa$, the initial energy of an electron $\varepsilon_0$ does not matter for the ``angle from bend'' $u\theta_0$ of recoil electron with whatever momentum transfer. 
In a sense, eq.~(\ref{noenergy}) is similar to the formula (1) from ref.~\cite{telnov_beam_2003} which describes the angle of spin $\theta_s$ with respect to the direction of motion of a relativistic electron:
\begin{equation}
\theta_s = \frac{\mu'}{\mu_0}\frac{\int \!\! B_\perp dl}{mc/e},
\label{spin}
\end{equation}
where $\mu'$ and $\mu_0$ are the anomalous and normal parts of the electron magnetic moment.

It is worth noting that there is the established method for the beam energy determination based on direct measurement of the maximal photon energy $\hbar\omega_{max}$ from table~\ref{scattering_parameters}, the details could be found in review~\cite{achasov_cms_2020} and references therein.

\subsection{Cross section}

ICS cross section depends on polarization states of all initial and final particles.
In case of averaging over the polarizations of the final states the cross section depends solely from the initial photon and electron polarizations.
Stokes parameters $\xi_1, \xi_2, \xi_3$ describe the polarization of laser light as it is explained in Appendix~\ref{Stokes}.
The electron beam polarization has three components $\zeta_x$, $\zeta_y$, $\zeta_z$, the total degree of polarization $\sqrt{\zeta^2_x+\zeta^2_y+\zeta^2_z}\in[0:1]$. 
We take differential cross section from ref.~\cite{berestetskii_quantum_1982} and after Lorentz transformations it is represented in $u$ and $\varphi$ variables by the sum of the six terms: 
\begin{equation}
\begin{aligned}
\frac{1}{r_e^2}\frac{d\sigma_0}{du\,d\varphi} & =  \frac{1}{\kappa(1+u)^3}\left[ 1 + (1+u)^2 - 
 4\frac{u}{\kappa}\left(1-\frac{u}{\kappa}\right)(1+u) \right],\\
\frac{1}{r_e^2}\frac{d\sigma_{\xi_1}}{du\,d\varphi} & =  \frac{4\,\xi_1}{\kappa(1+u)^2}\frac{u}{\kappa}\left(1-\frac{u}{\kappa}\right) \cos(2\varphi),\\
\frac{1}{r_e^2}\frac{d\sigma_{\xi_2}}{du\,d\varphi} & =  \frac{4\,\xi_2}{\kappa(1+u)^2}\frac{u}{\kappa}\left(1-\frac{u}{\kappa}\right) \sin(2\varphi),\\
\frac{1}{r_e^2}\frac{d\sigma_x}{du\,d\varphi} & =  \frac{-2\,\xi_3\,\zeta_x}{(1+u)^3} \; \frac{u}{\kappa}\sqrt{\frac{u}{\kappa}\left(1-\frac{u}{\kappa}\right)} \cos(\varphi),\\
\frac{1}{r_e^2}\frac{d\sigma_y}{du\,d\varphi} & =  \frac{-2\,\xi_3\,\zeta_y}{(1+u)^3} \; \frac{u}{\kappa}\sqrt{\frac{u}{\kappa}\left(1-\frac{u}{\kappa}\right)} \sin(\varphi),\\
\frac{1}{r_e^2}\frac{d\sigma_z}{du\,d\varphi} & =  \frac{\xi_3\,\zeta_z}{(1+u)^3} \; \frac{u}{\kappa}\left(u+2\right)\left(1-2\frac{u}{\kappa}\right).
\end{aligned}
\label{suall}
\end{equation}
In (\ref{suall}) $r_e$ is the classical electron radius and $d\sigma_0$ represents the cross section averaged over all polarization terms.
Integration of $d\sigma_0$ gives the total unpolarized cross section:
\begin{equation}
\sigma_0  =
\frac{2 \pi r_e^2}{\kappa} \left[ \left( 1-\frac{4}{\kappa}-\frac{8}{\kappa^2}\right) \log(1+\kappa)\, 
 +  \frac{1}{2}\left( 1-\frac{1}{( 1+\kappa)^2}\right) +\frac{8}{\kappa}\right].
\label{tcs}
\end{equation}
If $\kappa \ll 1$ (\ref{tcs}) tends to Thomson cross section $\sigma_0=\frac{8}{3}\pi r_e^2 \left( 1-\kappa\right)$.
The last term in (\ref{suall}) could modify the cross section given by (\ref{tcs}) in case $\xi_3\,\zeta_z \neq 0$.
Its integration gives the result:
\begin{equation}
\sigma_z=\xi_3\,\zeta_z\frac{2\pi r_e^2}{\kappa}\left[\left(1+\frac{2}{\kappa}\right)\log(1+\kappa)-\frac{1}{2}\left(4+\left(\frac{\kappa}{1+\kappa}\right)^2\right)\right].
\label{tcsz}
\end{equation}

\subsection{Photons}

The cross section with respect to the photon scattering angles $\eta_x$ and $\eta_y$ from table~\ref{scattering_parameters} could be derived from eqs.~(\ref{suall}) using the following transformations:
\begin{equation}
u          = \frac{\kappa}{1+\gamma^2\eta_x^2+\gamma^2\eta_y^2}, \;\;\;\; du d\varphi  =  \frac{2u^2\gamma^2}{\kappa} d\eta_x d\eta_y,
\label{ududndn}
\end{equation}
where we take $\theta_0=0$ for brevity.
The parts of the cross section are:
\begin{equation}
\begin{aligned}
\frac{1}{r_e^2}\frac{d\sigma_0}{d\eta_x d\eta_y} & =  \frac{2}{(1+u)^3} \left(\frac{\gamma u}{\kappa}\right)^2 \left[ 1 + (1+u)^2 - 
 4\frac{u}{\kappa}\left(1-\frac{u}{\kappa}\right)(1+u) \right],\\
\frac{1}{r_e^2}\frac{d\sigma_{\xi_1}}{d\eta_x d\eta_y} & =  \frac{8\,\xi_1}{(1+u)^2} \left(\frac{\gamma u}{\kappa}\right)^4 \left(\eta_x^2 - \eta_y^2\right),\\
\frac{1}{r_e^2}\frac{d\sigma_{\xi_2}}{d\eta_x d\eta_y} & =  \frac{16\,\xi_2}{(1+u)^2} \left(\frac{\gamma u}{\kappa}\right)^4 \eta_x\, \eta_y,\\
\frac{1}{r_e^2}\frac{d\sigma_x}{d\eta_x d\eta_y} & =  \frac{-4\,\xi_3\,\zeta_x}{(1+u)^3} \left(\frac{\gamma u}{\kappa}\right)^3 u\, \eta_x,\\
\frac{1}{r_e^2}\frac{d\sigma_y}{d\eta_x d\eta_y} & =  \frac{-4\,\xi_3\,\zeta_y}{(1+u)^3} \left(\frac{\gamma u}{\kappa}\right)^3 u\, \eta_y,\\
\frac{1}{r_e^2}\frac{d\sigma_z}{d\eta_x d\eta_y} & =  \frac{2\xi_3\,\zeta_z}{\gamma(1+u)^3} \left(\frac{\gamma u}{\kappa}\right)^3 (u+2)(\kappa-2u).
\end{aligned}
\label{xsgamma}
\end{equation}
Equations (\ref{xsgamma}) are applicable for describing the spatial distribution of scattered photons measured e.~g. by pixel detector with rectangular geometry.

\subsection{Electrons}
To make the formulae in this section more compact and easy to use we introduce modified variables 
\begin{equation}
\vartheta_x = \gamma\theta_x,\;
\vartheta_y = \gamma\theta_y,\;
\vartheta_0 = \gamma\theta_0.
\end{equation}
From definitions given in table~\ref{scattering_parameters} we obtain the equation:
\begin{equation}
(\vartheta_x - u\vartheta_0)^2 + \vartheta_y^2 = u(\kappa-u),
\label{nequality}
\end{equation}
whose roots are
\begin{equation}
u_\pm=\frac{\kappa/2+\vartheta_0\vartheta_x\pm\sqrt{\kappa^2/4-\vartheta_x^2-\vartheta_y^2(1+\vartheta_0^2)+\kappa\vartheta_0\vartheta_x}}{1+\vartheta_0^2}.
\label{upm}
\end{equation}
In the $\vartheta_x,\vartheta_y$ plane all recoil electrons are enclosed within the ellipse described by the discriminant of eq.~(\ref{upm}).
The center of this ellipse is located at the point $[\vartheta_x =\kappa\vartheta_0/2; \vartheta_y=0]$, its vertical semi-minor axis equals $\kappa/2$ while horizontal semi-major axis is $\sqrt{1+\vartheta_0^2}$ times longer.
The transformation from $u,\varphi$ to $\vartheta_x,\vartheta_y$ variables is specified by the Jacobian matrix:

\begin{equation}
\begin{bmatrix}
\displaystyle\frac{\partial\vartheta_x}{\partial u} & \displaystyle\frac{\partial\vartheta_x}{\partial\varphi}\\[1em]
\displaystyle\frac{\partial\vartheta_y}{\partial u} & \displaystyle\frac{\partial\vartheta_y}{\partial\varphi}
\end{bmatrix}
=
\begin{bmatrix*}[r]
\displaystyle \vartheta_0 - \frac{\kappa/2-u}{\sqrt{u(\kappa-u)}}\cos\varphi & &
\displaystyle \sqrt{u(\kappa-u)}\sin\varphi \\[1em]
\displaystyle -\frac{\kappa/2-u}{\sqrt{u(\kappa-u)}}\sin\varphi & &
\displaystyle -\sqrt{u(\kappa-u)}\cos\varphi
\end{bmatrix*}.
\label{nJacob}
\end{equation}
The Jacobian determinant:
\begin{equation}
\begin{aligned}
\det (J(u,\varphi))
& = & \kappa/2 - u - \vartheta_0\sqrt{u(\kappa-u)}\cos\varphi = \\
= \det(J(\vartheta_x,\vartheta_y)) & = &  \pm \sqrt{\kappa^2/4-\vartheta_x^2-\vartheta_y^2(1+\vartheta_0^2)+\kappa\vartheta_0\vartheta_x} .
\end{aligned}
\label{DJ}
\end{equation}
$D = |\det(J(\vartheta_x,\vartheta_y))|$ equals the absolute value of the square root of discriminant in eq.~(\ref{upm}). 
The following expressions follow from the definitions given in table~\ref{scattering_parameters}:
\begin{equation}
\label{stxty}
\begin{aligned}
\cos(\varphi)  &= \frac{u\vartheta_0-\vartheta_x}{\sqrt{u(\kappa-u)}},& \;& \sin(\varphi)          &=&  \frac{-\vartheta_y}{\sqrt{u(\kappa-u)}}, \\
\cos(2\varphi) &= \frac{(u\vartheta_0-\vartheta_x)^2-\vartheta_y^2}{u(\kappa-u)},& \;& \sin(2\varphi) &=&  \frac{-2(u\vartheta_0-\vartheta_x)\vartheta_y}{u(\kappa-u)}.
\end{aligned}
\end{equation} 
Lets apply the variables transformation to eqs.~(\ref{suall}):
\begin{equation}
\begin{aligned}
\frac{1}{r_e^2}\frac{d\sigma_0}{d\vartheta_xd\vartheta_y} & = \frac{1}{D\kappa(1+u)^3}\left[ 1 + (1+u)^2 - 4\frac{u(\kappa-u)}{\kappa^2}(1+u) \right],\\
\frac{1}{r_e^2}\frac{d\sigma_{\xi_1}}{d\vartheta_xd\vartheta_y} & =  \frac{ 4\xi_1 }{D\kappa^3(1+u)^2}\left((u\vartheta_0-\vartheta_x)^2-\vartheta_y^2\right),\\
\frac{1}{r_e^2}\frac{d\sigma_{\xi_2}}{d\vartheta_xd\vartheta_y} & =  \frac{-8\xi_2 }{D\kappa^3(1+u)^2}(u\vartheta_0-\vartheta_x)\vartheta_y,\\
\frac{1}{r_e^2}\frac{d\sigma_x}{d\vartheta_xd\vartheta_y} & =  \frac{-2\xi_3\,\zeta_x }{D\kappa^2(1+u)^3} u(u\vartheta_0-\vartheta_x),\\
\frac{1}{r_e^2}\frac{d\sigma_y}{d\vartheta_xd\vartheta_y} & =  \frac{ 2\xi_3\,\zeta_y }{D\kappa^2(1+u)^3} u\vartheta_y,\\
\frac{1}{r_e^2}\frac{d\sigma_z}{d\vartheta_xd\vartheta_y} & =  \frac{\xi_3\,\zeta_z }{D\kappa^2(1+u)^3} \; u(u+2)(\kappa-2u).
\end{aligned}
\label{txtyall}
\end{equation}
The total cross section is defined by the sum of these terms, each of which is calculated twice (for $u_+$ and for $u_-$).
With the following linear transformation instead of the ellipse we obtain the circle of unit radius with the center located at the origin of new coordinate system:
\begin{equation}
\begin{aligned}
x =  A\frac{\vartheta_x}{\kappa/2} - B; \hspace{5mm} y = \frac{\vartheta_y}{\kappa/2}, \text{ where}\\
A=\frac{1}{\sqrt{1+\vartheta_0^2}} \text{ and } B = \frac{\vartheta_0}{\sqrt{1+\vartheta_0^2}}.
\end{aligned}
\label{xy}
\end{equation}
The condition $\vartheta_0 \gg 1$ obviously provides better separation of recoil electrons from the beam and in this case $A\simeq 1/\vartheta_0$ and $B\simeq 1$.
Exact calculations give the following:
\begin{equation}
\begin{aligned}
\frac{d\vartheta_x d\vartheta_y}{D}  &=  \frac{\kappa\,dx\,dy}{2\sqrt{1 - x^2 - y^2}} ,\\
u_\pm                  &=  \frac{\kappa}{2}\left( 1 + B x \pm A\sqrt{1-x^2-y^2}\right) = \frac{\kappa}{2}\left( 1 + \Delta_\pm\right),\\
(u\vartheta_0-\vartheta_x)_\pm &= \frac{\kappa}{2}\left( \;\; -\, A x \pm B\sqrt{1-x^2-y^2}\right) = \frac{\kappa}{2}\delta_\pm.
\end{aligned}
\label{janduxy}
\end{equation}
Parameters $\Delta^\pm$ and $\delta^\pm$ in eqs.~(\ref{janduxy}) are the shorthand's for corresponding parenthesized expressions.
We see that $d\vartheta_x d\vartheta_y/D$ provides the cross section ``amplification'' near the border of the circle, while the difference $u_+-u_-$ is at maximum in the center of the circle.
Finally we obtain:
\begin{equation}
\begin{aligned}
\left.\frac{1}{r_e^2}\frac{d\sigma_0}{dxdy}\right|_\pm       & = &               &\frac{1 + (1+u_\pm)^2 - (1+u_\pm)(1-\Delta_\pm^2)}{2(1+u_\pm)^3\sqrt{1-x^2-y^2}},\\
\left.\frac{1}{r_e^2}\frac{d\sigma_{\xi_1}}{dxdy}\right|_\pm & = & \xi_1         &\frac{\delta_\pm^2-y^2}{2(1+u_\pm)^2\sqrt{1-x^2-y^2}},\\
\left.\frac{1}{r_e^2}\frac{d\sigma_{\xi_2}}{dxdy}\right|_\pm & = & \xi_2         &\frac{-\delta_\pm y}{(1+u_\pm)^2\sqrt{1-x^2-y^2}},\\
\left.\frac{1}{r_e^2}\frac{d\sigma_x}{dxdy}\right|_\pm       & = & \xi_3 \zeta_x &\frac{-u_\pm\delta_\pm }{2(1+u_\pm)^3\sqrt{1-x^2-y^2}},\\
\left.\frac{1}{r_e^2}\frac{d\sigma_y}{dxdy}\right|_\pm       & = & \xi_3 \zeta_y &\frac{u_\pm y }{2(1+u_\pm)^3\sqrt{1-x^2-y^2}},\\
\left.\frac{1}{r_e^2}\frac{d\sigma_z}{dxdy}\right|_\pm       & = & \xi_3 \zeta_z &\frac{-u_\pm(u_\pm+2)\Delta_\pm}{2(1+u_\pm)^3\sqrt{1-x^2-y^2}} .
\end{aligned}
\label{xse}
\end{equation}
The cross section in $x,y$ plane is the sum of ``+'' and ``-'' solutions.
It's time to explain why the variables $x$ and $y$ were introduced. 
To build a function that describes the number of events in a pixel of the detector of recoil electrons, we need the integral of the cross section over a rectangular area.
The antiderivative of $f(x,y) = 1/\sqrt{1-x^2-y^2}$ function is an analytical expression
\begin{equation} 
F(x,y) = x\atan2(y,\rho) + y\atan2(x,\rho) - \atan2(xy,\rho),
\label{anti}
\end{equation} 
where $\rho = \sqrt{1-x^2-y^2}$. 
The integral over the pixel with the corners $x_0,y_0$ and $x_1,y_1$  is 
\begin{equation} 
I(x_0,y_0,x_1,y_1) = F(x_0,y_0) + F(x_1,y_1) - F(x_0,y_1) - F(x_1,y_0).
\label{inte}
\end{equation} 

\section{Monte Carlo experiment}
\subsection{General considerations}
Consider again the diagram shown in figure~\ref{fig:cxema}.
In the Monte-Carlo experiment scattered photons and electrons hit the notional detectors located far enough from the interaction area and the bending magnet.
These detectors are installed along the $x$-axis on the right-side of figure~\ref{fig:cxema}. 
It is also worth paying attention that there are no magnetic elements between the magnet and the detectors.

If $R_0$ is the bending radius of a beam electron, then the radius of the recoil electron with minimal energy equals $R_\text{min}=R_0/(1+\kappa)$.
Consider the bending dipole of length $L$ with rectangular shape of the magnetic poles.
All the electrons enter the dipole along the same line -- the beam orbit.
The equality $\Delta\theta=\kappa\theta_0$ in table \ref{scattering_parameters} was obtained under the assumption that the field integral does not depend on the electron energy.
The accuracy of this assumption depends on the parameters combination $\Lambda = \kappa L /2 R_0$. 
At the end of the dipole an electron with minimal energy $\varepsilon_{min}$ will have transverse horizontal distance from the beam orbit $\Delta x \simeq \Lambda L$, 
which sets the requirement for a transverse size of the uniform field region inside the magnet.
Relative difference in the arcs lengths inside the magnet for $\varepsilon_0$ and $\varepsilon_{min}$ electrons is estimated as $\Delta L / L \simeq \Lambda^2/2$.
With $\Lambda\simeq10^{-3}$ the relative difference in the magnetic field integrals $\Delta L / L \simeq 5 \cdot 10^{-7}$.
Small enough values of $\Lambda$ are accessible with small bending angles $L/R_0 \ll 1$ when $\kappa \geq 1$.
At low electron beam energies, when $\kappa \ll 1$, the bending angle could be increased.

\begin{figure}[h]
\centering
\includegraphics[width=0.65\linewidth]{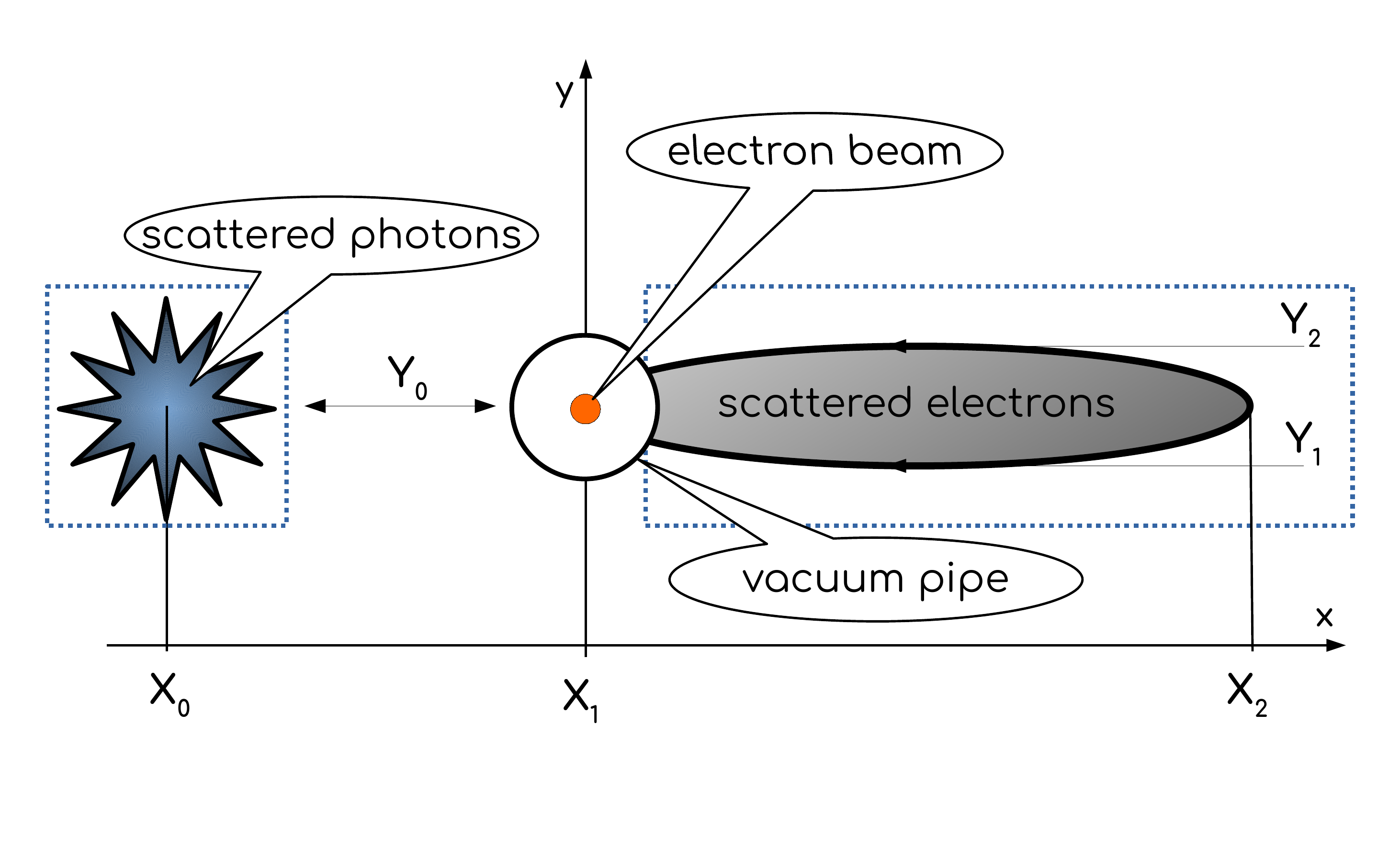}
\caption{Detection plane. 2D pixel detectors for photons and electrons are represented by dotted rectangles.}
\label{fig:principle}
\end{figure}

Use of two-dimensional pixel detector for recoil electrons was proposed for transverse beam polarization measurement in ref.~\cite{mordechai_transverse_2010}.
Figure~\ref{fig:principle} illustrates the scattered photons, recoil electrons and corresponding detectors in $x,y$ plane.

It is assumed that the detectors are precisely positioned relative to each other.
Each detector has a certain geometry, number of pixels and pixel size in both dimensions.
In the numerical experiment the detectors are represented by the two-dimensional ($x,y$) histograms where each bin corresponds to one pixel.
The following parameters are used in further Monte-Carlo simulations:

\begin{itemize}
\item $\gamma$ is the mean value of the beam electron Lorentz factor.
\item $\theta_0$ is the equilibrium orbit bending angle.
\item $\lambda_0$ and $\hbar\omega_0=hc/\lambda_0$ are the wavelength and energy of a laser photon.
\item $\kappa=4\gamma\hbar\omega_0/mc^2$ is the mean value of the Compton scattering parameter $\kappa$.
\item Stokes vector $[\xi_1, \xi_2, \xi_3]$ describes the laser beam polarization.
\item Vector $[\zeta_x, \zeta_y,\zeta_z]$ describes the electron beam polarization.
\item $u$ and $\varphi$ parameters are generated according to the sum of eqs.~(\ref{suall}).
\item Lorentz factor of a beam electron is $\gamma' = \gamma(1+\Delta)$  where $\Delta$ is extracted from the normal distribution with zero mean value and standard deviation described by the relative energy spread of the electron beam $\sigma_\gamma/\gamma$. 
\item The transverse horizontal coordinate of the electron at the moment of scattering due to synchrotron motion is $D_x \Delta $ where $D_x$ is the dispersion function. 
\item The transverse coordinates $x,y$ of the electron due to betatron motion are generated by the normal distributions with zero mean value and $\sigma_x=\sqrt{\epsilon_x \beta_x}$, $\sigma_y=\sqrt{\epsilon_y \beta_y}$ described by corresponding emittance and $\beta$-function.
\item The transverse angles $x',y'$ of the electron due to betatron motion are generated by the normal distributions with zero mean value and $\sigma_x'=\sqrt{\epsilon_x/ \beta_x}$, $\sigma_y'=\sqrt{\epsilon_y/\beta_y}$.
\item The length of the interaction region of laser radiation with electron beam is assumed to be very short ($\ll L_1$) and thus it is not taken into account. 
\item The $X_\gamma, Y_\gamma, X_e, Y_e$ coordinates of scattered photon and electron at the detection plane are obtained by the following equations:
\end{itemize}
\begin{equation}
\begin{aligned}
X_\gamma & = \frac{1}{\cos\!\theta_0}\left(D_x\Delta  + x + L_1\tan\!\left[x' + \frac{\cos\!\varphi\sqrt{\kappa/u-1}}{\gamma(1+\Delta)}\right]\right)
- L_2\tan\!\theta_0 , \\
Y_\gamma & = \hspace{11mm} y + L_1\tan\left[y'+ \frac{\sin\!\varphi\sqrt{\kappa/u-1}}{\gamma(1+\Delta)}\right], \\
X_e      & = \frac{1}{\cos\!\theta_0}\left(D_x\Delta  + x + L_1\tan\!\left[x' - \frac{\cos\!\varphi\sqrt{u(\kappa-u)}}{\gamma}\right]\right)
 + L_2 \tan\!\theta_0 \left[u-\frac{\Delta}{1+\Delta}\right], \\
Y_e      & = \hspace{11mm} y + L_1\tan\left[y'- \frac{\sin\!\varphi\sqrt{u(\kappa-u)}}{\gamma}\right].
\label{mc}
\end{aligned}
\end{equation}

If the photon $X_\gamma, Y_\gamma$ or electron $X_e, Y_e$ coordinates hits the pixel of the photon/electron detector, the content of the corresponding histogram bin is increased by one. 
Note that the eqs.~(\ref{mc}) contain the lattice functions $\beta_x, \beta_y, D_x$ but not their derivatives.
Each beam electron at the area of interaction with laser radiation has some deviations from the equilibrium parameters due to betatron and synchrotron motion.
The transverse coordinates of this electron at the plane of the detectors installation could be obtained from the full set of its initial parameters by tracking through the spectrometer.
According to eq.~(\ref{noenergy}) bending of the recoil electrons differs from bending of the beam electrons by an additional angle which does not depend on the initial electron parameters.
It means that the (gaussian) blur of the two-dimensional distribution of recoil electrons described by eqs.~(\ref{xse}) is equal to the transverse beam size ($\sigma_x, \sigma_y$) at the plane of detectors.

It is worth noting that in this simulation it is implicitly assumed that the scattering probability does not depend on the transverse coordinates of the electron.
This condition is satisfied if the size of the laser spot significantly exceeds the size of the electron beam.
Such a configuration of the interaction area is necessary in any case when we are going to measure the true average parameters of the electron beam.

\subsection{FCC-ee beam polarimeter}

FCC-ee is a lepton collider with centre-of-mass energies between 90 and 350 GeV~\cite{abada_fcc-ee_2019}.
It is considered as a potential intermediate step towards the realization of the hadron facility. 
Beam energy calibration by resonant depolarization is the basis for the precise measurements of the $Z$ mass and width with a precision of <100~keV, and of the $W$ mass and width with a precision of the order of 500~keV.
The conceptual design of the Compton polarimeter may be found in~\cite{abada_fcc-ee_2019, muchnoi_performance_2019} and the same setup is considered here.
\begin{table}[h]
\caption{Simulation parameters. All designations have been defined in the text above.}
\label{MCPAR}
\centering
\begin{tabular}{|l|r|ll|l|l|}
\hline
$\varepsilon_0=\;45.6$~GeV & $\gamma=$~89237 & $\epsilon_x=270$ & pm rad & $\beta_x=100$~m  & $L_1=117$~m  \\
\hline
$\lambda_0=\;532$     ~nm  &  $\kappa=1.6279$ & $\epsilon_y=1$ &pm rad &  $\beta_y=\;\,20$~m & $L_2=100$~m \\
\hline
$\omega_0=2.331$    ~eV  & $\vartheta_0=$~190.44 & $\theta_0=2.1341$ & mrad & $D_x=\,25$~mm & $\sigma_\gamma/\gamma=$~0.001\\
\hline
\end{tabular}
\end{table}
Table~\ref{MCPAR} contains the list of simulation parameters which will be identical in all numerical experiments below.
Bunch revolution frequency at FCC-ee is $3 \cdot 10^3\;\text{s}^{-1}$ while the rate of Compton scattering events is estimated as $2 \cdot 10^6\;\text{s}^{-1}$.
The parameters of the pixel detectors for photons and electrons are presented in Table~\ref{DPAR}.
\begin{table}[h]
\caption{Detectors: geometry, number of pixels, size of pixels.}
\label{DPAR}
\centering
\begin{tabular}{|l|l|l|l|l|}
\hline
Detector  & Size ($X \times Y$) & $N$pix ($X \times Y$) &  Pixel size ($X \times Y$) \\
\hline
Photons   & $10 \times 10$~mm   &  $100 \times 100$     &  $100 \times 100\;\mu$m  \\
\hline
Electrons & $400 \times 4$~mm   & $1600 \times 80$      &  $250 \times 50\;\mu$m \\
\hline
\end{tabular}
\end{table}

\subsubsection{Experiment I}

The goal of the first numerical experiment is to verify the above formulae for the cross sections and check the fitting procedure.
The subject of this study is the trivial case when the Stokes vector of laser polarization is $[\xi_1, \xi_2, \xi_3] = [0,0,1]$ and the electron beam is unpolarized $\zeta_x, \zeta_y, \zeta_z = [0,0,0]$.
\begin{figure}[h]
\centering
\vspace{5mm}
\includegraphics[width=0.5\linewidth]{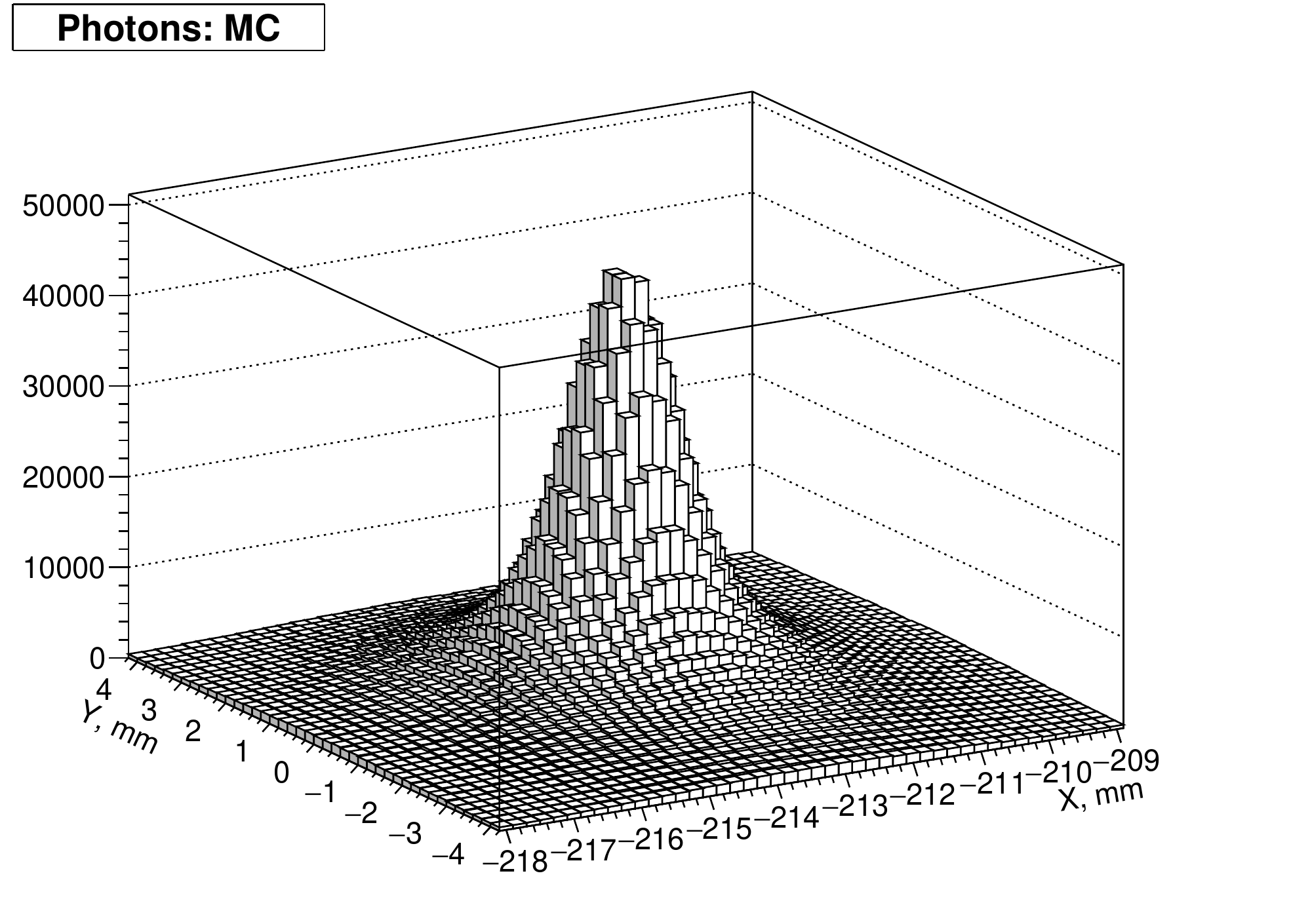}
\caption{Photon spot at the detector. Number of bins is reduced for better visualization.}
\label{fig:P1}
\end{figure}
Figure~\ref{fig:P1} shows the distribution of scattered photons at the detector obtained with $10^7$ Monte-Carlo events.
In FCC-ee conditions~\cite{muchnoi_performance_2019} this amount is available within 5~s measurement time with approximately $1.5\cdot10^{4}$ collisions between laser pulse and electron bunch.  

The fit function for the scattered photons distribution is based on eqs.~(\ref{xsgamma}) with nine fit parameters $X_0$, $Y_0$, $\sigma_x$, $\sigma_y$, $\xi_1$, $\xi_2$, 
$\xi_3\zeta_x$, $\xi_3\zeta_y$, $\xi_3\zeta_z$ representing the spot center position, blur effect from electron beam emittance and five independent polarization components.
The tenth parameter is the normalization.
The distance from the interaction area to the detector $L_1$ and the scattering parameter $\kappa$ are fixed according to the setup.
The results of the fit are presented in Table~\ref{FITP1} (NDF is the acronym for Number of Degrees of Freedom).
\begin{table}[h]
\caption{Photon spot fit results.}
\label{FITP1}
\centering
\begin{tabular}{|l|l|}
\hline
$X_0 = -213.538 \pm 0.001$~mm    & $Y_0 = -0.002 \pm 0.001\;$mm \\
$\sigma_x =255 \pm 3 \;\mu$m     & $\sigma_y = 30 \pm 18 \;\mu$m \\
$\xi_1 = 0.000 \pm 0.002$        & $\xi_2 = -0.001 \pm 0.001 $ \\
$\xi_3\zeta_x = 0.004 \pm 0.006$ & $\xi_3\zeta_y = -0.008 \pm 0.006$ \\
$\xi_3\zeta_z = 0.000 \pm 0.002$ & $\chi^2/\text{NDF}=9796.9/9990$ \\
\hline
\end{tabular}
\end{table}

It can be seen that the position of the spot is established with accuracy about 1~$\mu$m.
The $\sigma_x$ and $\sigma_y$ values carry the information about the combination of beam parameters at the interaction area.
The accuracy of $\sigma_y$ is poor because it is three times smaller than the pixel size.
It would be useful to recall that in order to detect gamma rays, they will be converted into $e^+e^-$ pairs, which will lead to additional blurring and reduce the number of registered events.
However all polarization parameters correspond to their zero set values with absolute accuracy of better than one percent.

\begin{figure}[h]
\centering
\includegraphics[width=0.5\linewidth]{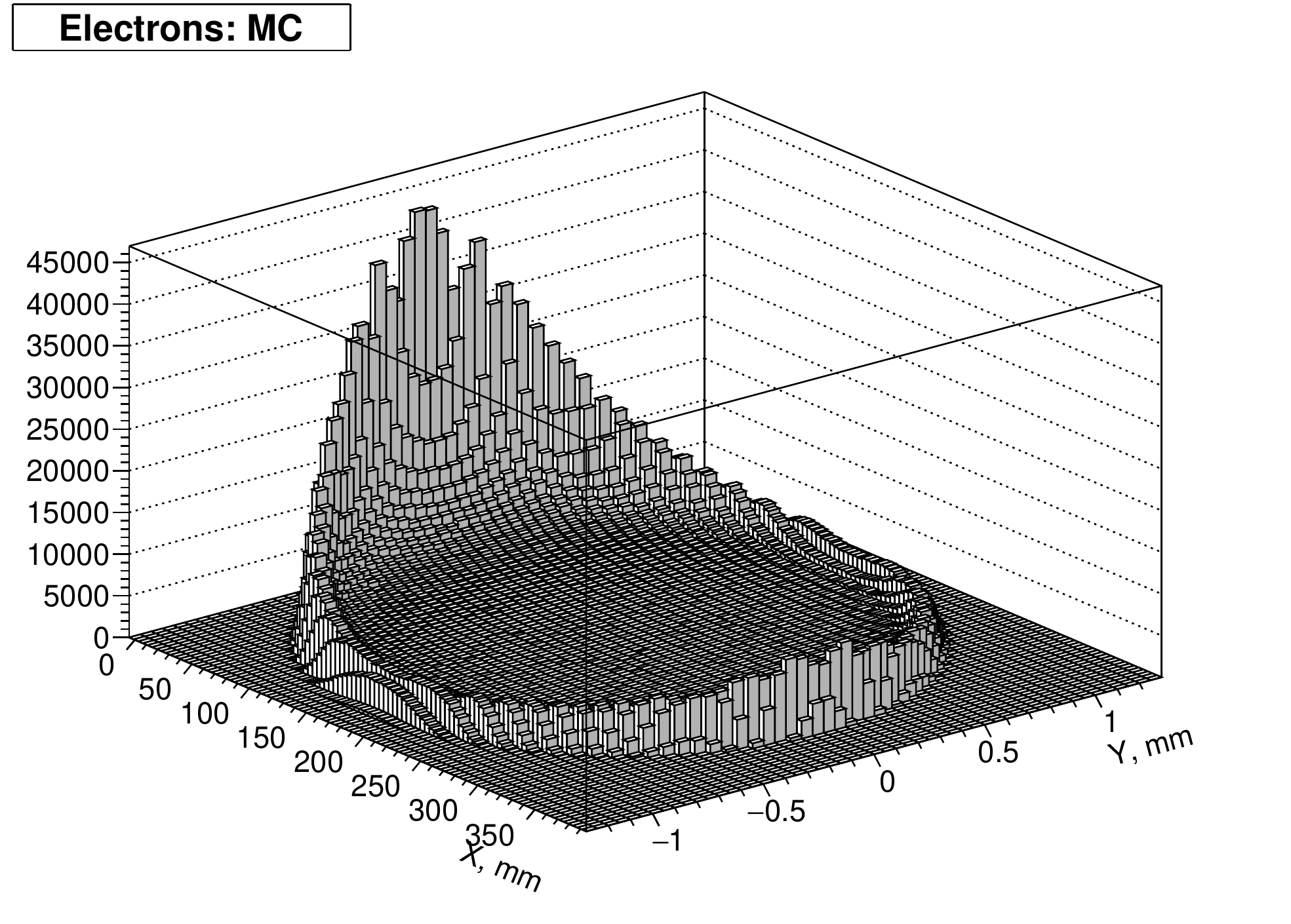}
\caption{Recoil electrons at the detector. Number of $x$-bins is reduced to 100 for better visualization. Electrons detector covers the beam at the coordinates $X=0, Y=0$, which is impossible to do in practice.}
\label{fig:E1}
\end{figure}

Let us take advantage of the opportunities provided by the numerical experiment and place the electron detector so that it detects all recoil electrons, including those that propagate very close to the beam.
Such a distribution of scattered electrons is presented in figure~\ref{fig:E1}.

The fit procedure is done by Pearson's method, the expected bin error is estimated as $\sqrt{1+\mathcal{F}}$, where the fit function $\mathcal{F}(X,Y)$ is calculated in three steps.
First, geometry parameters $X_1$, $X_2$, $Y_1$, $Y_2$ shown in figure~\ref{fig:principle} specify ellipse of scattered electrons defined in eqs.~(\ref{upm}) and (\ref{DJ}).
With linear transformations, based on the known detector geometry, these parameters are converted to dimensionless variables $x$ and $y$ defined in eqs.~(\ref{xy}).
The integrals from (\ref{inte}) are calculated as well as the corresponding cross section terms from eqs.~(\ref{xse}) at the center of each pixel.
Scattering parameter $\kappa$ is fixed as it was done with photons while neither $L_1$ nor $L_2$ are not used.
Finally the distribution, obtained by pixel by pixel multiplication of the integrals and the cross sections, is convoluted with two-dimensional Gaussian to account for transverse beam sizes at the detection plane.
Totally the fit function has twelve parameters including normalization, while only non-empty bins are accounted in NDF.
The results of the fit are presented in table~\ref{FITE1}.
\begin{table}[h]
\caption{Scattered electrons ellipse fit results.}
\label{FITE1}
\centering
\begin{tabular}{|l|l|}
\hline
$X_1          =  0.0035 \pm 0.0016$~mm   & $X_2 = 347.635 \pm 0.003$~mm \\
$Y_1          = -1.0682 \pm 0.0001$~mm   & $Y_2 = 1.0682 \pm 0.0001$~mm \\
$\sigma_x     = 320.7   \pm 1.5 \;\mu$m  & $\sigma_y = 27.06 \pm 0.03 \;\mu$m \\
$\xi_1        = 0.001   \pm 0.001$       & $\xi_2 = 0.432 \pm 0.198 $ \\
$\xi_3\zeta_x = 1.000   \pm 0.195$       & $\xi_3\zeta_y = -0.001 \pm 0.002$ \\
$\xi_3\zeta_z = 0.000   \pm 0.001$       & $\chi^2/\text{NDF}=51568.9/52270$ \\
\hline
\multicolumn{2}{|c|}{$\varepsilon_0 = 45.5997\pm0.0007$~GeV}\\
\hline
\end{tabular}
\end{table}

It can be seen that the accuracy of the left ellipse border position $X_1$ is 0.16~$\mu$m which is approximately twice better than the accuracy of the right border position $X_2$.
This is explained by larger amount of recoil electrons near $X_1$ (the beam position).
The difference $Y_2-Y_1$ is defined with relative precision of $5\cdot10^{-5}$, which allows to measure (or just check) $L_1$ with the same accuracy.
The $X_1$ and $(Y_1+Y_2)/2$ define the electron beam transverse position relative to the detector.
Unlike the photons distribution, the $\sigma_x$ and $\sigma_y$ here are the parameters describing the transverse beam sizes at the detector position and we see that both sizes could be measured with rather high accuracy (compared to e.~g. ref.~\cite{honda_upgraded_2005}).
As for polarization parameters, $\xi_2$ and $\xi_3 \zeta_x$ determination is wrong.
This happens because the sum of $u_+$ and $u_-$ solutions in corresponding terms of the cross section (\ref{xse}) have opposite signs and almost cancel each other when $\vartheta_0 \gg 1$ (in our case $\vartheta_0 = \gamma\theta_0\simeq 190$).
The other three polarization parameters $\xi_1$, $\xi_3 \zeta_y$ and $\xi_3 \zeta_z$ however are determined even more precise than from the photons distribution.
The value of the beam energy $\varepsilon_0$ in the last row of table~\ref{FITE1} is obtained from $X_0, X_1, X_2$ fit parameters with relative accuracy of about $1.5\cdot10^{-5}$ by the formula which follows from eq.~(\ref{noenergy}):
\begin{equation}
\label{beam_energy}
\varepsilon_0 = \frac{(mc^2)^2}{4\omega_0}\frac{X_2-X_1}{X_1-X_0}.
\end{equation}
 
\subsubsection{Experiment II} 

In this experiment we study more realistic case when the recoil electron detector is shifted 15~mm away from the electron beam.
It is assumed that this gap will provide a sufficient physical aperture for the electron beam.
The Stokes vector of laser polarization is chosen as $[\xi_1, \xi_2, \xi_3] = [0.1, 0.1, 0.99]$ with small amount of vestigial linear polarization which may exist due to imperfect polarization control.
The set of the electron beam polarization parameters is $[\zeta_x, \zeta_y, \zeta_z] = [0.1, 0.25, 0.1]$.
The expected polarization of the pilot electron bunches at FCC-ee (averaged over thousands beam revolutions) has $\zeta_y$ component only.
The $\zeta_x$ and $\zeta_z$ components are added to the simulations in order to investigate the possibility of measuring the electron beam polarization in general case.

We do not present the figure with the distribution of photons here since visually it is very close to that shown in figure~\ref{fig:P1}.
The fitting procedure is the same as in previous experiment and the results obtained from the fit are presented in Table~\ref{FITP2}. 
From the table we can conclude that all polarization parameters are determined correctly with absolute accuracies from 0.1\% to 0.7\%.
\begin{table}[h]
\caption{Photon spot fit results.}
\label{FITP2}
\centering
\begin{tabular}{|l|l|}
\hline
$X_0 = -213.539 \pm 0.002$~mm    & $Y_0 = 0.000 \pm 0.001\;$mm \\
$\sigma_x = 246 \pm 4 \;\mu$m    & $\sigma_y = 13 \pm 70 \;\mu$m \\
$\xi_1 = 0.102 \pm 0.002$        & $\xi_2 = 0.100 \pm 0.001 $ \\
$\xi_3\zeta_x = 0.095 \pm 0.007$ & $\xi_3\zeta_y = 0.247 \pm 0.006$ \\
$\xi_3\zeta_z = 0.105 \pm 0.002$ & $\chi^2/\text{NDF}=9935.8/9990$ \\
\hline
\end{tabular}
\end{table}

\begin{figure}[h]
\centering
\includegraphics[width=0.5\linewidth]{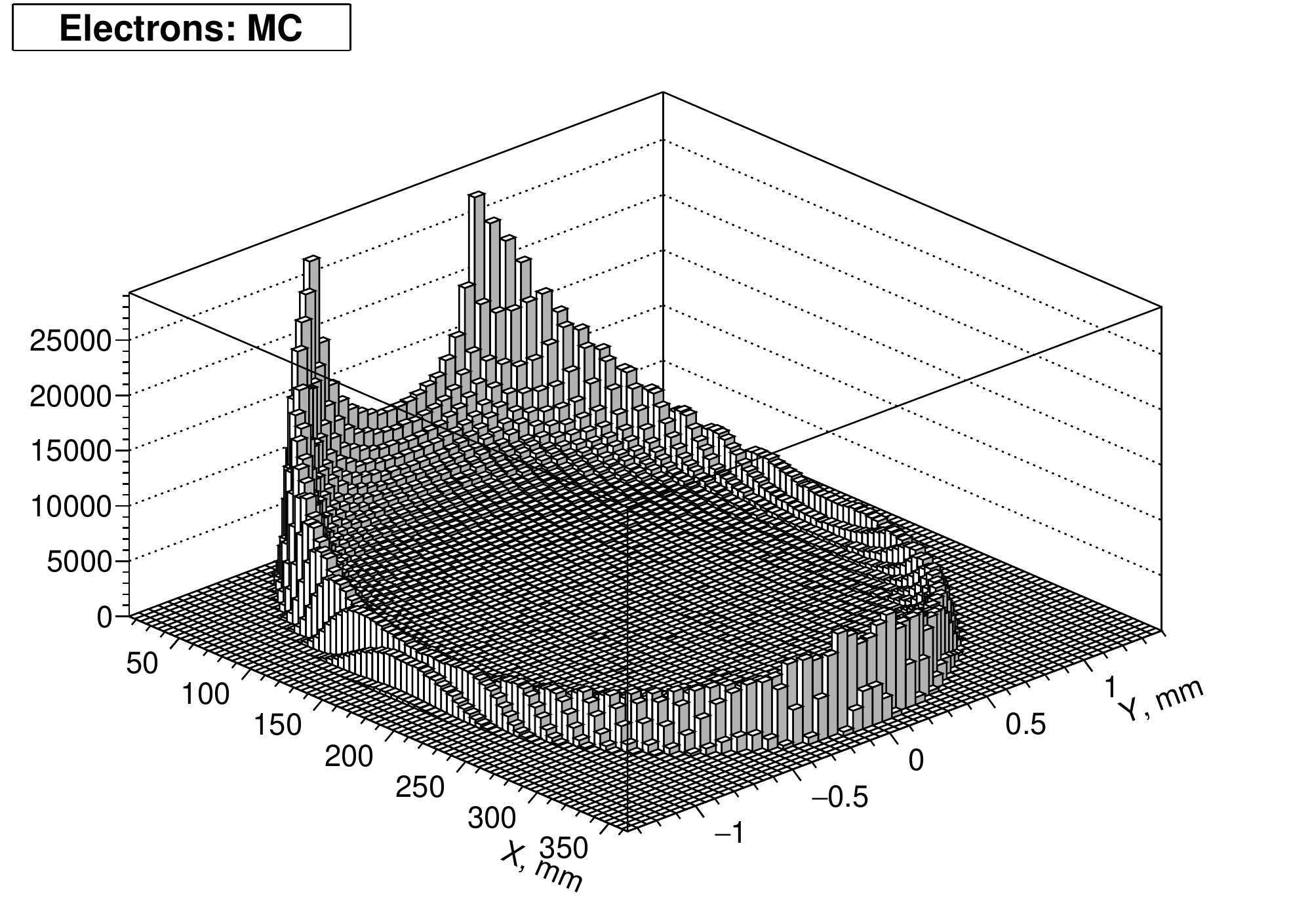}
\caption{Recoil electrons at the detector. Number of $x$-bins is reduced to 100 for better visualization. Electrons detector is shifted in $X$ by 15~mm away from the beam  coordinates $X=0, Y=0$.}
\label{fig:E2}
\end{figure}

The distribution of recoil electrons in figure~\ref{fig:E2} looks different mainly due to the new location of the detector,
the results of the fit are presented in table~\ref{FITE2}. 

\begin{table}[h]
\caption{Scattered electrons ellipse fit results.}
\label{FITE2}
\centering
\begin{tabular}{|l|l|}
\hline
$X_1          =  0.013 \pm 0.007$~mm   & $X_2 = 347.632 \pm 0.004$~mm \\
$Y_1          = -1.0682 \pm 0.0001$~mm   & $Y_2 = 1.0684 \pm 0.0001$~mm \\
$\sigma_x     = 319.6   \pm 4.3 \;\mu$m  & $\sigma_y = 27.15 \pm 0.03 \;\mu$m \\
$\xi_1        = 0.100   \pm 0.001$       & $\xi_2 = 0.100 $ \\
$\xi_3\zeta_x = 0.099            $       & $\xi_3\zeta_y = 0.246 \pm 0.002$ \\
$\xi_3\zeta_z = 0.099   \pm 0.001$       & $\chi^2/\text{NDF}=50152.7/51245$ \\
\hline
\multicolumn{2}{|c|}{$\varepsilon_0 = 45.5959\pm0.0025$~GeV}\\
\hline
\end{tabular}
\end{table}
The fitting was performed in the same way as in our previous experiment, except that the parameters $\xi_2$ and $\xi_3 \zeta_x$ are fixed according to the results obtained from the photons distribution. 
We see that the accuracy of determining parameters $X_1$ and $X_2$ is almost the same now.
The transverse beam sizes $\sigma_x$ and $\sigma_y$ are in perfect agreement with the results of the first experiment.
The polarization parameters $\xi_1$, $\xi_3 \zeta_y$ and $\xi_3 \zeta_z$ are determined with absolute accuracy of 0.1\%, 0.2\% and 0.1\%.
The beam energy $\varepsilon_0$ is ``measured'' with relative accuracy of about $5.5\cdot10^{-5}$ and differs from the true value by a little less than two standard deviations.
The accuracy is worse than in the first experiment cause the electrons detector is shifted away from the electron beam by 15~mm and $X_1$ parameter is now determined by extrapolation of the shape of distribution.

\section{Summary}
Simultaneous measurement of the spatial distributions of gamma quanta and recoil electrons makes it possible to significantly expand the capabilities of the Compton polarimeter.
In addition to all polarization components of the laser and electron beams, another parameters can be measured with rather high accuracy: the average energy of beam electrons, the integral of the field in the spectrometer magnet, the transverse coordinates and dimensions of the electron beam at the location of the detectors.
All these parameters are obtained by analysis of the two-dimensional spatial distribution of recoil electrons.
The recoil electrons have a sharp boundary in the scattering angle which, after bending in the magnet, takes the form of an ellipse in the detection plane. 
The sharp edges of the ellipse are blurred by the gaussian beam sizes of the electron beam at the location of the detectors.

The considered approach is applicable when the vertical size of the ellipse of recoil electrons is larger than the vertical beam size:
\begin{equation}
L_1 \frac{2 \omega_0}{mc^2} > \sqrt{\varepsilon_y\beta_y}.
\end{equation}
Recoil electrons propagate to the inner side of the beam orbit and thus there is no direct background from high energy synchrotron radiation.
The angular distribution of scattered photons partially overlaps with photons of synchrotron radiation, this circumstance requires additional consideration if we are dealing with high electron energies.
Electrons are registered directly by ionization losses while photons should be converted into pairs leading to a decrease in the detection efficiency and spatial resolution.
Despite the fact that the fluxes of scattered photons and electrons are almost the same, the flux density of electrons is much lower due to bending and corresponding spatial separation by energies, making easier the simultaneous detection of multiple recoil electrons.

The main technical difficulty for the practical implementation of the proposed measurement scheme is the need to have a long spectrometer arm, in which the recoil electrons are not exposed to magnetic fields. 
A high degree of field uniformity in the bending magnet is required for accurate determination of the beam energy. 
Additional simulations with realistic parameters of the magnetic system should be carried out to estimate the systematic effects due to these factors. 
If the realistic situation differs from the ideal one, the successful fitting of the recoil electron distributions can be performed by introducing additional fit parameters to the transformation from the detector coordinates $X$ and $Y$ (figure \ref{fig:principle}) to the dimensionless coordinates $x$ and $y$ used in eqs.~(\ref{xse}).

The FCC-ee polarimeter was designed to provide fast and precise measurements of the transverse polarization of the pilot bunches.
This will allow to continuously monitor the beam energy by resonant depolarization technique at the beam energies up to 80~GeV approximately, when it becomes impossible to keep the beam polarized.
For these needs the analysis of the distribution of recoil electrons will provide more accurate measurements at the same acquisition time if compared with scattered photons.
Meanwhile the energy spectrometer option of the polarimeter could be cross-checked with the resonant depolarization results and allow to use it at higher beam energies and for interpolation purposes.
Detection and analysis of distributions of both photons and recoil electrons will reduce systematic errors in measurements of all polarization components.
Accurate determination of the transverse sizes of the electron beam is, in a sense, a usefull additional bonus.

\appendix
\section{Stokes parameters}\label{Stokes}
The Stokes parameters describe the polarization state of electromagnetic radiation. 
Their definition is slightly different in different sources, so below are the definition that we use here.
\begin{itemize}
\label{poldefs}
\item $\xi_0 = E_x^{\,2}+E_y^{\,2}$ is the intensity of light. \\
With normalization $E_x^{\,2}+E_y^{\,2}=1$ for 100\% polarized laser radiation $\xi_0=\sqrt{\xi_1^2+\xi_2^2+\xi_3^2}=1$.
\item $\xi_1 = E_x^{\,2}-E_y^{\,2}$.
\begin{itemize}
\item[] $E_x=1, E_y=0, \xi_1=+1$: \hfill 100\% linear polarization along x-axis. 
\item[] $E_x=0, E_y=1, \xi_1=-1$: \hfill 100\% linear polarization along y-axis.
\end{itemize}
\item $\xi_2 = 2 E_x E_y\cos(\delta)$. 
\begin{itemize}
\item[] $E_x=E_y, \delta=0, \xi_2=+1$: \hfill 100\% linear polarization along $\varphi=+\pi/4.$ 
\item[] $E_x=E_y, \delta=\pi, \;\xi_2=-1$: \hfill 100\% linear polarization along $\varphi=-\pi/4.$ 
\end{itemize}
\item $\xi_3 = 2 E_x E_y\sin(\delta)$. 
\begin{itemize}
\item[] $E_x=E_y, \delta=+\pi/2, \;\xi_3=+1$: \hfill 100\% right circular polarization. 
\item[] $E_x=E_y, \delta=-\pi/2, \;\xi_3=-1$: \hfill 100\% left circular polarization.
\end{itemize}
\end{itemize}

\acknowledgments

The author expresses his gratitude to Mikhail Achasov and Ivan Koop for useful discussions, to Alain Blondel and Eugene Levichev for their interest and motivation to continue these studies. 
The author would also like to thank Katsunobu Oide who always found a solution to the difficult issue of polarimeter integration to the FCC-ee lattice (otherwise these studies would have been closed in the early stages of discussion) and Aur\'elian Martens for his help in deriving the cross section formulae.
This study is dedicated to the upcoming centennial anniversary of the discovery of the Compton effect.
\bibliographystyle{JHEP}
\bibliography{elbepoensp}

\end{document}